# A Comparison of Information Retrieval Techniques for Detecting Source Code Plagiarism

Vasishtha Sriram Jayapati
College of Information and Computer Sciences
University of Massachusetts Amherst
Amherst, MA, USA
vjayapati@cs.umass.edu

Ajay Venkitaraman
College of Information and Computer Sciences
University of Massachusetts Amherst
Amherst, MA, USA
avenkitarama@cs.umass.edu

## ABSTRACT
Plagiarism is a commonly encountered problem in the academia. While there are several tools and techniques to efficiently determine plagiarism in text, the same cannot be said about source code plagiarism. To make the existing systems more efficient, we use several information retrieval techniques to find the similarity between source code files written in Java. We later use JPlag, which is a string-based plagiarism detection tool used in academia to match the plagiarized source codes [1]. In this paper, we aim to generalize on the efficiency and effectiveness of detecting plagiarism using different information retrieval models rather than using just string manipulation algorithms.

## INTRODUCTION
Plagiarism is an issue that plagues many areas of academia. The most common forms of plagiarism that we see are in the homework reports and submissions that students make during their academic career. In the field of computer science, another form of plagiarism is very much prevalent, and that is the plagiarism in programming assignments, commonly referred to as source code plagiarism. The amount of resources that are at the disposal of students with the proliferation of online source codes makes this issue even more pronounced. The detection of source code plagiarism is different (and in a way, more difficult) from that of other types of plagiarism because of the following reasons:

- Understanding the intent of source code is more difficult than understanding the intent of prose, because in many problems, there are only a finite number of solutions.
- Source code can be intelligently modified so as not to resemble the original code and there is more than one way to accomplish this.
- Many programming assignments have starter codes, and boilerplate code which causes a basic level of similarity between source codes, in turn leading to difficulty in the estimation of extent of plagiarism, whereas in prose it is more straightforward to understand the extent of plagiarism.

Previous work in this include some renowned plagiarism detection systems that are primarily based on string manipulation such as JPlag [1], MOSS [2] and SHERLOCK [3]. Even though these systems are effective and are widely used in various universities to detect plagiarism in programming assignments, they employ string manipulation algorithms (which cannot run in linear time complexity) over all the input files, thus slowing down the process of plagiarism detection considerably in case of large code repositories.

On the other hand, information retrieval techniques have been used to retrieve relevant documents for a given query in an efficient manner. Our system proposes to use these information retrieval techniques to fetch a set of relevant codes for every given source code and then uses JPlag to scan this set of codes for plagiarism.

## RELATED WORK
Source code plagiarism detection has been a well discussed problem over the last couple of decades and great strides have been taken in the effective detection of plagiarism. MOSS is one of the most popular systems that deals with plagiarism detection over 20 languages including but not limited to C, C++, C#, Java, Python, JavaScript, FORTRAN, etc. [9]. MOSS uses robust form of winnowing, which is a document fingerprinting algorithm and it outputs the results in the form of HTML pages which shows the key areas of similarity between documents [2]. JPlag is another plagiarism detection engine which is implemented in Java, where the input is a directory containing the source codes to be inspected and the output is a directory of HTML pages which shows the details of the similarity between the source codes. The user interface is descriptive, and it shows a list of the plagiarized source codes according to highest similarity and average similarity. The system tokenizes the source codes and uses Greedy String Tiling algorithm to compare the tokenized source codes [1]. SHERLOCK uses incremental comparison to compare the source codes five times after various modifications. The source codes are tokenized as a part of the preprocessing during one of the comparisons. The output is a neural network that is represented in the form of a PostScript image which illustrates similarities between the various source codes [3].

## DATASET

We have used the dataset provided by FIRE (Forum for Information Retrieval Evaluation) which is available on-demand. This dataset consists of the training corpus with 259 source codes in Java along with their relevance judgements for tuning the hyperparameters of the retrieval models. It also contains 12,080 source codes in the test corpus which have been used to evaluate our system [4].

## METHODOLOGY

The system that we propose tries to reduce the time required to find most of the plagiarized source code pairs in an efficient manner using information retrieval techniques. Firstly, we define precision and recall, which are two metrics that are commonly used in the paper.

$$precision = \frac{no.\,of\,plagiarized\,source\,code\,pairs\,retreived}{no.\,of\,retrieved\,source\,code\,pairs}$$

$$recall = \frac{no.\,of\,plagiarized\,source\,code\,pairs\,retreived}{total\,no.\,of\,plagiarized\,source\,code\,pairs}$$

The two key objectives of our system are to get a high value of recall, and to try and retrieve as few documents as possible for a high recall, which will ensure that the system is highly efficient. The flowchart (Figure 1) below explains the working of the system.

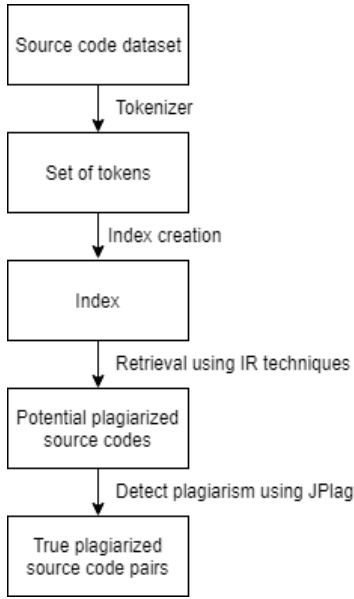

**Figure 1: Flowchart of the working of the system**

Initially, we have the dataset mentioned above with the source codes, some of which are plagiarized. We preprocess the data and convert the source code into tokens. Once the tokens are created, we use Galago search engine [5] to create the index. This index is then queried by the search engine to detect potential plagiarized source code pairs. We have experimented with three information retrieval techniques, viz. Okapi BM25 [6], Query Likelihood Model using Dirichlet Prior Smoothing [7] and Probabilistic Model based on Divergence from Randomness [8] to find the best possible retrieval technique for our dataset. We briefly describe the three techniques below.

Okapi BM25 model is based on the BM25 weighting scheme and it is a non-binary, probabilistic model which is sensitive to document length and term frequency. The relevance score for a query document pair is given by the following equation:

$$\sum_{t \in q \cap d} \frac{(k+1) * tf(t,d)}{k\left(1 - b + b\frac{|d|}{avgl}\right) + tf(t,d)} \log \frac{N - df(t) + 0.5}{df(t) + 0.5}$$

Where $b$ (typically considered as 0.75) and $k$ (typically considered as 1.2) are constants, and $tf(t,d)$ is the term frequency of term $t$ in document $d$ (whose length is given by $|d|$). The number of documents in which the term $t$ has appeared is given by $df(t)$. The average length of all documents in the corpus is denoted by $avgl$ and the total number of documents in the corpus is denoted by $N$ [6].

Query Likelihood Model using Dirichlet Prior Smoothing is a Bayesian smoothing model which uses the Dirichlet Distribution prior which accounts for the document length. The probability of a term given a document model is given by:

$$p_\mu(w|d) = \frac{c(w;d) + \mu p(w|C)}{\sum_w c(w;d) + \mu}$$

Where $\mu$ is a constant, $p(w|C)$ is the probability of the word given the corpus model and $c(w;d)$ is the count of the word $w$ in the document $d$ [7].

Probabilistic Model based on Divergence from Randomness (PDFR) is as described in [8]. The normalized term frequency is given as:

$$tf_n = tf.\log\left(1 + \frac{avg\_l}{l}\right)$$

Where $avg\_l$ is the average length of the document in the corpus and $l$ is the length of the document observed [8].

We note here that for our specific problem, both the query and the document are source codes which are to be analyzed. We use the open source implementation of these models implemented as a part of the Galago search engine [5]. We check the precision and recall values by retrieving a range of potentially plagiarized codes for every given source code. This list is ranked based on the similarity measure of the source code as measured by the above models. On the training data we compare the various models and their precision and recall values which will decide the best model out of the three and the final hyperparameters for plagiarism detection on the test data. As will be shown in the evaluations, the best model for the given data was Okapi BM25. We then use JPlag to compute the accuracy of the model on the test data, by detecting the true plagiarized source code pairs. Here we assume the results of JPlag to be the ground truth, which is a fair assumption, since JPlag is one of the most accurate plagiarism detection systems (which is also well maintained as an open source project).

## EVALUATION

We wish to evaluate the system on the following two measures:
- The accuracy of the system, which we measure using precision and recall (as defined in the methodology above).
- The time taken to detect the plagiarized source code pairs.

Here we compute and plot the precision vs recall curve on the training data for the different information retrieval models (as discussed in the methodology), the time taken to detect plagiarism for different values of recall on the training data, the time taken to detect plagiarism for different number of Top-N retrieved source code pairs on the test data using the best performing model and finally measuring the accuracy of our predictions about the plagiarized source codes on the test data. All these evaluations were performed on a laptop with 4 GB 1600 MHz DDR3 RAM, 2.6 GHz Intel Core i5-3230M processor running Ubuntu 18.10.

Figure 2 below shows the precision vs recall curve for the different information retrieval models on the training data. It is quite evident from the figure that the state-of-the-art Okapi BM25 retrieval function provides the best precision for all the recall values as compared to the other methods. Also, the Query Likelihood Model using Dirichlet Prior Smoothing seems to fetch more potentially plagiarized pairs (thereby increasing precision) for lower values of recall as compared to PDFR model. We also see that PDFR outperforms the Query Likelihood Model as the recall tends towards 1.

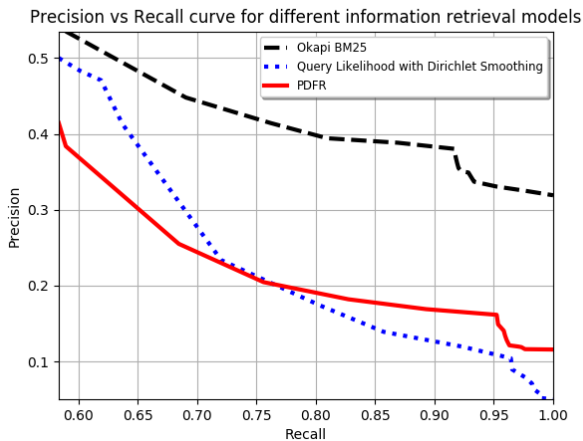

Figure 2: Precision vs Recall curve

Figure 3 shows the time taken to evaluate plagiarism for different values of recall on the training data. Having a recall value of 1 implies that we consider all the pairs of plagiarized codes. For a recall value of 1, the best performance was given by Okapi BM25 model, taking about 1.59 seconds for plagiarism evaluation. This includes the time taken by the search engine to retrieve the list of Top-N potential plagiarized codes for every source code and the time taken to run JPlag on these retrieved source code pairs. Evaluating plagiarism on the entire set of source code pairs in the training data using JPlag took about 6.48 seconds. Therefore, using the information retrieval similarity technique as a pre-filter provided a speedup of about 400%. Also, it can be observed that different retrieval models perform differently at various recall values, so the models can be chosen based on the tradeoff between speedup and the possibility of losing out on some plagiarized pairs.

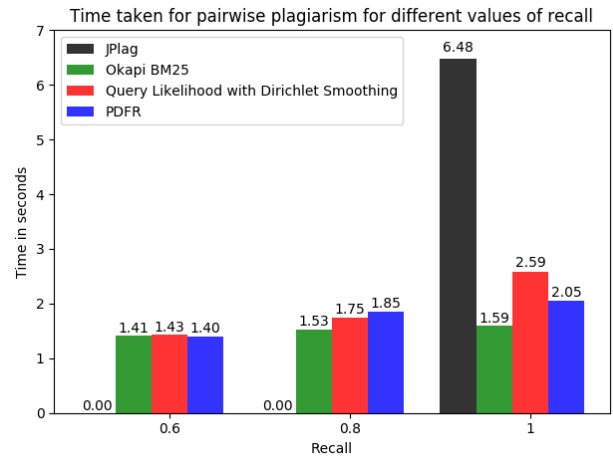

Figure 3: Plagiarism Detection Time vs Recall

Figure 4 illustrates the time taken to evaluate plagiarism for different number of Top-N retrieved source code pairs on the both train and test data using the best performing model, which was Okapi BM25 in our scenario. The time taken by JPlag to scan all the pairs of codes for plagiarism was 6.48 seconds and 3 minutes and 59 seconds on train and test data respectively. Hence, we can see that the speedup provided on the test data is 2180%, which is what we expected, owing to the larger size of the test dataset as compared to the train dataset. This provides an understanding on the impact of using information retrieval techniques as filter before using the existing plagiarism detection tools.

The evaluation of the accuracy of the best retrieval model (Okapi BM25) on the test data is provided in Table 1. We define accuracy as follows:

$$accuracy = \frac{no.\,of\,plagiarized\,pairs\,retrieved\,by\,Okapi\,BM25}{no.\,of\,plagiarized\,pairs\,retrieved\,by\,JPlag}$$

| Top-N retrieved source codes | Accuracy (in percentage) |
|---|---|
| 2 | 71.67 |
| 5 | 88.33 |
| 12 | 96.67 |
| 16 | 100 |

Table 1: Accuracy values for different Top-N retrieved codes

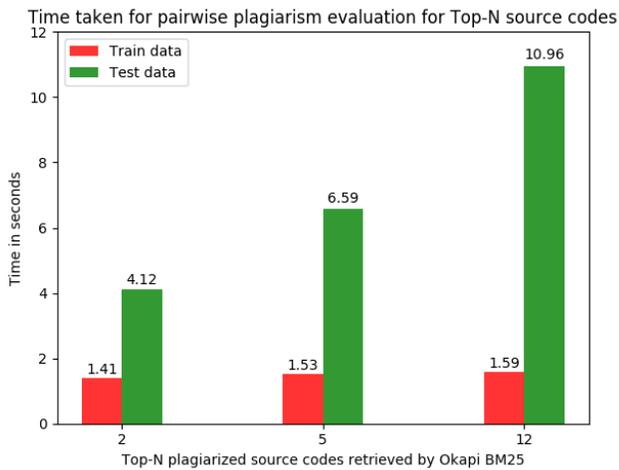

**Figure 4: Plagiarism Detection Time vs Top-N retrieved source codes**

## CONCLUSION

In this paper we implemented a highly efficient plagiarism detection system that scales linearly with the number of documents in the corpus. Considering recall to be the primary measure of accuracy, it is possible to tune the hyperparameters in such a way that the system can be more accurate or less accurate with a tradeoff with respect to the time taken for plagiarism detection. Based on our evaluation results, we find that the Okapi BM25 model performed the best for the given dataset as compared to the other retrieval models discussed. But it is important to note that the best model depends on the dataset and hence it is imperative to tune the hyperparameters which provides the best results for the particular dataset.

## FUTURE WORK

As an extension to our current research, it will be interesting to see how we can use different tokenization approaches such as using a bytecode approach which can detect most introductory-programming-course plagiarism attacks at any level by utilizing low-level instructions instead of source code tokens [10]. We can also evaluate the system using several other similarity measures, as proposed in [11], [12]. Future work can also delve into different thresholding mechanisms [13] to filter the retrieved source code pairs using the information retrieval models.